\documentstyle[preprint,aps]{revtex}
\begin{document}
\draft

\title{Weakly Chaotic Population Dynamics in Random Ecological Networks}

\author{Shin-ichi Sasa}

\address{Department of Physics, Kyoto University, Kyoto 606, Japan
\footnote{ Address as of  Aug. 16, 1994:
Department of Pure and Applied Sciences,
         College of Arts and Sciences, University of Tokyo,
         Komaba, Meguro-ku, Tokyo 153, Japan}}

\author{Tsuyoshi Chawanya}
\address{Yukawa Institute for Theoretical Physics,  Kyoto 606, Japan}

\date{August 10, 1994}
\maketitle

\begin{abstract}
Population dynamics in random ecological networks
are investigated by analyzing a  simple deterministic equation.
It is found that a  sequence of abrupt changes of populations
punctuating quiescent states  characterize  the long time behavior.
An asymptotic analysis  is developed by introducing a log-scaled time,
and it is shown  that  such a dynamical process behaves as  non-steady
weak chaos in which population disturbances grow algebraically in time.
Also, some relevance of our study to taxonomic data of biological
extinction is mentioned.
\end{abstract}




Almost all possible equilibrium  populations in random
ecological networks become  unstable
when the number of  fully connected species is
sufficiently large \cite{Gardner,May,Robert}.
This is a general result obtained from
random matrix theory \cite{Gardner}.
The loss of  stability of equilibrium populations, of course,
dose not necessarily imply   a catastrophe of the system;
it is known that limit cycles,  heteroclinic cycles, and  chaos
arise  beyond the loss of the stability in  a few species system
\cite{Schuster,May2,Hofbauer,Maynard}.
Therefore, it seems natural to consider
how populations evolve  in  ecological systems consisting
of  many randomly interacting species.


This  problem is  related to the  co-evolution of many species
in an unchanging physical environment.
Each species evolves  in the effective environment formed by the others,
and the evolution of one species  alters the environment of many others.
This   leads to  a kind of frustration in the ensemble of species,
and if the frustration is never resolved by the evolution,
it would continue  indefinitely.
Such a scenario was proposed by Van Valen \cite{Valen} and
is often referred as the Red Queen hypothesis \cite{Maynard,Valen}.

Our  aim is to present significant features of population
dynamics in  a random ecological network by analyzing  suitable  mathematical
models. The  choice of  models  may be a controversial step, and
various types of  dynamical behavior will  be observed depending on
this choice. We postpone  discussion  about this step,
and in this paper  we focus on a specific model which is composed of
simple deterministic equations for the populations of $N$ species.
The equations are so simple that we can employ an asymptotic
analysis for a long time behavior of the populations.



Our  equations  describe the  time development of populations
$\{x_i\}$ of $N$ species.
We assume that each species  grows  at  a rate  $\lambda_i$
which is expressed  by a Lotka-Volterra type coupling
with the other species:
\begin{equation}
\lambda_i=a_i+\sum_j g_{ij} x_j,
\label{eqn:LV}
\end{equation}
where $a_i$ and  $g_{ij}$  correspond to an intrinsic growth rate
and competing coefficients, respectively.
We further assume that each population has a saturation level, and we
normalize the populations so that $x_i=1$ at the  saturation levels.
$x_i$ then obeys the  logistic equation:
\begin{equation}
{d x_i \over d t}
=\lambda_i(x_i-x_i^2),
\label{eqn:logi}
\end{equation}
where $0 \le x_i \le  1$, and
$x_i=0$ and $x_i= 1$  correspond to  'extinction' and 'saturation',
respectively.
Although  we consider the case that $a_i=0$  in  (\ref{eqn:LV})
for simplicity, the  analysis given below is easily extended so as to
include  this term.  {}From our motivation, $N$ is a given large number,
and $g_{ij}$ is  assumed to be a random variable obeying
a Gaussian distribution with  zero mean and deviation $J/\sqrt{N}$
except for $g_{ii}=0$.
Note that  the value of $J$ can be renormalized to 1
by changing the  time scale.


Equation (\ref{eqn:logi}) has $2^{N}$ equilibrium states,
each of which is expressed by
$(x_1, x_2,\cdots, x_N)=(\sigma_1, \sigma_2, \cdots \sigma_N)$,
where $\sigma_i \in \{0,1\}$.  In an equilibrium state
$(\sigma_1, \sigma_2, \cdots, \sigma_N)$,
species satisfying $\sigma_i=1$  exist at their saturation levels,
while other species are all extinct.
We call the  special state with  $x_i=0$ for all $i$ 'perfect extinction'.
The linear stability of equilibrium states is determined  by
the eigenvalues of the linearized equation around these  states.
It can be easily shown that the probability of choosing
a  stable state from  all  equilibrium states equals $1/2^{N}$,
and therefore almost all equilibrium states correspond to  saddle points.
Although  the expectation  of  the number of stable equilibrium
states is approximately 1, whether or not a given  system can reach
such a state depends on the  global structure of the flow  in phase space.


We  first  give results of numerical simulations
of (\ref{eqn:LV}) and (\ref{eqn:logi}) in Fig. \ref{fig1}.
It is seen that a population abruptly  increases or decreases
from a  quiescent state  close to  an equilibrium.
We call the  quiescent state and abrupt change
'quasi-equilibrium' and 'burst', respectively.
Such  dynamical behavior seems to continue indefinitely
within our computationally available time.
In fact, even if the system eventually settles into a stable  equilibrium,
a long transient time is needed to  reach the neighborhood of
this state.
We  can estimate the waiting time as $ O[ \exp ( 2^{N}) ]$,
because the system needs to experience $O(2^N)$ bursts to meet
the stable equilibrium state, and the time interval between two successive
bursts  grows exponentially on the  average.
(The latter statement will be demonstrated in the argument below.)
We thus focus on the long time behavior
composed of 'quasi-equilibrium' and 'burst'
irrespective of the final state.


Now, we develop a new asymptotic method.
Our key idea is to introduce a new variable $y_i$
defined as to satisfy
\begin{equation}
x_i=f(y_i):={\exp(y_i) \over 1+\exp(y_i)}.
\label{eqn:trans}
\end{equation}
Then, Equation (\ref{eqn:logi}) is transformed into a simple form:
\begin{equation}
{d y_i \over d t}=\lambda_i.
\label{eqn:yeq}
\end{equation}
Integrating this equation, we obtain
\begin{equation}
y_i(t)=y_i(0)+<\lambda_i>t,
\end{equation}
where $<\lambda_i>$  is the average of the growth rate
over the time interval from 0 to $t$ and is written as
\begin{equation}
<\lambda_i>={1 \over t} \int_0^{t} dt^{\prime} \lambda_i(t^\prime).
\label{eqn:deflambda}
\end{equation}
Since $<\lambda_i>$  does not converges  to zero  for $ t \rightarrow \infty$
except in  the case that the system approaches  the perfect extinction,
we obtain the asymptotic form of $x_i$ for $t \rightarrow \infty$:
\begin{equation}
x_i(t)=f(y_i(0)+<\lambda_i>t) \rightarrow \theta(<\lambda_i>).
\label{eqn:longx}
\end{equation}
That is, after a sufficiently long time,
the system stays at  a quasi-equilibrium state, and
whether $x_i$ is close to $0$ or to  $1$
is determined by the sign of $<\lambda_i>$.  We say, in this sense,
that  the dynamics of the population is slaved to that of
the average growth rate. We thus consider  time development of
$<\lambda_i>$.
The definition of $<\lambda_i>$ given by (\ref{eqn:deflambda}) leads
immediately to the equation:
\begin{equation}
{d <\lambda_i> \over d t}=
{ 1\over t}(- <\lambda_i>+ \sum_j g_{ij} x_j),
\end{equation}
where we have used the expression of the growth rate (\ref{eqn:LV}).
Then, by introducing  a log-scaled time $\tau=\log(t)$,
defining a new variable  $h_i(\tau)=<\lambda_i>$, and  using
the asymptotic form (\ref{eqn:longx}),
we obtain the equation describing the slow dynamics of
the average growth rate:
\begin{equation}
{d h_i \over d \tau}=
- h_i+\sum_j g_{ij}\theta(h_j).
\label{eqn:somp}
\end{equation}
Here, it is worth noting that the equations of motion
for  the average growth rates are  autonomous
when we use a log-scaled time $\tau$.
We now consider the relationship  between  the  orbits
of $h_i(\tau)$ and
$x_i(t)$. First, from the asymptotic form (\ref{eqn:longx}),
$x_i$ is close to 0 or 1 depending on the sign of $h_i$, and
a burst of the $i$-th species is identified with a zero-crossing
of $h_i$. Second, consider periodic motion of $h_i$ expressed by
$h_i(\tau)=\sin( 2 \pi \tau / T)$.
Then, a log-scaled time at which $h_i$ crosses  zero
for the $n$-th time is given by $\tau_n=Tn/2 $.  Correspondingly,
zeros of $ <\lambda_i> $ occur at  $\exp (Tn/2)$. This means that
the time interval between two successive bursts grows exponentially in $n$.
We can see in general that  a periodic orbit of $h_i(\tau)$ corresponds to a
transient orbit of $x_i(t)$ attracting to a heteroclinic cycle.
Finally, the long time behavior of populations corresponding to
chaotic motion of $h_i$ is  described by a non-steady process
of an irregular occurrence of bursts.
The  time interval between two successive bursts  grows
irregularly in $n$, but on the average grows exponentially.

We also  note that Equation (\ref{eqn:somp})
has a form similar to the random neural network model
proposed by Sompolinsky et.al. \cite{Sompolinsky}  which
is obtained when we replace $\theta(h_j)$ by $\tanh(K h_j)$.
They studied  the statistical properties of fluctuations
by developing a new method called 'dynamical mean field theory'
and showed that their equation  exhibits chaotic behavior
when $ KJ >1 $ in the limit $N \rightarrow \infty$
\cite{Sompolinsky}.
Therefore, one may  guess  that
(\ref{eqn:somp}) has chaotic solutions for sufficiently large $N$.



In order to investigate dynamical properties  of  solutions  $h_i(\tau)$,
we  performed  numerical simulations of (\ref{eqn:somp}).
Noting that the equation is  piece-wise linear
and discontinuous at $h_i=0$, we can construct a solution
starting from an initial condition $\{h_i(0)\}$
in the following way.
First, the equation (\ref{eqn:somp}) can be integrated
analytically  until  a time $\tau_{1}$
at which the first  burst occurs, and $h_i(\tau_1)$ is expressed by
\begin{equation}
h_i(\tau_1)=(h_i(0)-\lambda_i)\exp(-\tau_1)+\lambda_i,
\label{eqn:ht}
\end{equation}
where $\lambda_i=\sum_j g_{ij}\theta(h_j)$ takes
a constant value during this period, and
$\tau_{1}$ is given by
\begin{equation}
\tau_{1}= \hbox{ Min}\kern-10pt\lower5.0pt\hbox{$_i$}\quad
\log(1-{ h_{i}(0) \over \lambda_i }),
\end{equation}
where the index $i$  runs over the
species satisfying  $h_i(0) \lambda_i < 0 $.
By repeating a similar procedure, we obtain a sequence of $\tau_i$
with $(i=1,2,  \cdots)$,
 and simultaneously the  solution of (\ref{eqn:somp}).

In Fig. \ref{fig2}, we plotted the species which bursts at  $\tau_i$
with $(i=1,2,  \cdots)$.
The plotted pattern seems to be  random along the time axis,
but  certain inhomogeneity in species can be seen.
In fact, Fig. \ref{fig3} shows that the time interval
between two successive  bursts obeys  a  Poisson distribution
$P(\tau)$, while  as shown in Fig.\ \ref{fig4}, $Q(\tau)$,
the distribution of  persistence time of positive $h_i$,
has  a power law tail
\begin{equation}
Q(\tau) \sim \tau^{-2}
\label{eqn:power}
\end{equation}
for $\tau \rightarrow \infty$,
which is  due to the
randomness of the long-time average of $h_i$.
Therefore, in the random neural network model proposed by Sompolinsky  et.al.
\cite{Sompolinsky},  similar  behavior is never observed \cite{Sasa}.
{}From the forms  of these distributions, we can expect that
the dynamics of $h_i(\tau)$ are steady.
As a result, on the real time scale,
the population dynamics are non-steady;
the time interval between two successive bursts grows in
$n$ irregularly, but on the average exponentially.
Further we numerically calculated the  maximum Lyapunov number
$\mu$ for many set  $g_{ij}$ chosen randomly, and found this
number to be positivein each case.
Thus, a  disturbance  $\delta h_i(\tau)$ grows
as $\delta h_i(\tau) \sim \exp(\mu \tau)$, that is,
time series of $ h(\tau) $ are orbitally unstable.
This result leads to power law divergence of
a disturbance of  populations $\delta x_i(t)$
of the form
\begin{equation}
 \delta x_i(t) \sim t^{\mu},
\end{equation}
and thus zero Lyapunov exponents. Our population dynamics
therefore are  not chaotic, but  'weakly chaotic' \cite{Bak}.
Discussion of the distribution function of $\mu$
for the ensemble of $\{g_{ij}\}$, the Lyapunov spectrum,
their dependence  on $N$, and a proof of (\ref{eqn:power})
will be presented in a separate paper \cite{Sasa}.



We  have shown  the significant features of the long time behavior of
populations  described by (\ref{eqn:LV}) and (\ref{eqn:logi}).
These  are expected to be  common to a large class of
models of random ecological networks.
Recently, population dynamics in random ecological networks
have been studied  by analyzing another type of equation
\cite{Opper,Fontana}.
This equation is  obtained by normalizing
the sum of the populations to  unity, and the resultant form is
the same as the replicator model proposed in the context of
molecular evolution \cite{Schuster,Hofbauer}.
We have found that a non-steady process with a log-scaled time
appears generally in a random network \cite{Cha},
and one of the authors has investigated a transition to
'chaos with a log-scaled time' in a four dimensional system \cite{Cha2}.


Here, we  present a remark:
a small perturbation of the dynamical system,
e.g., the addition of $\epsilon F(x_i)$ to the right-handed side of
(\ref{eqn:logi}),  yields topologically unequivalent orbits $x_i(t)$.
In other words, the above solutions are  structurally unstable.
This  structural instability is  analogous
to   that of homoclinic orbits \cite{Guckenheimer}.
Since  structurally unstable behavior cannot be observed
without certain constraints, one may object that the behavior
we study is not generic. We believe however that an understanding of
the behavior of the unperturbed system  will provide
the first step toward understanding the rich variety
of dynamical phenomena of perturbative systems much as
center manifold theory has succeeded
in  describing  complex behavior near bifurcation points \cite{Guckenheimer}.


We finally discuss the relevance of our study to
the taxonomic data of  biological extinction.
According to Van Valen \cite{Valen},
the  lifetime of species obeys a Poisson distribution
with an extinction rate $\Omega$ which is about
$10^{-7}$ [year$^{-1}$].   (Such a simple picture has been
called 'continuous extinction'. The alternative to this is
the 'episodic extinction' picture \cite{Raup}. )
Since  extinction ($x_i=0$) never occurs in our model,
we need to introduce  another assumption in order to see
the correspondence with extinction in taxonomic data.
One  plausible idea is  to regard  an  abrupt decrease
of a  population from near the saturation level as
extinction.  This may be justified,
because the taxonomic data are made from  fossil records.
Then, noting that the time spent near the saturation level
measured with a log-scaled time corresponds to
the persistence time of positive $h_i$,
we find that the lifetime  of a species obeys the distribution
$Q(\tau)$. This result seems to be inconsistent with the taxonomic data,
because our lifetime is measured by a log-scaled time.
The apparent paradox is resolved by taking a small perturbation into account.
In fact, we can show that a  certain  class of perturbations  alters
the expression of the  scaled time  $\tau$ from  $\tau=\log( t)$  to
$\tau=\epsilon t$ while keeping the $\tau$ dependence of statistical
expressions unchanged \cite{Sasa}. $\epsilon$ here measures
the slowness of the time scale which is related to the magnitude
of the perturbation.
A similar change of a time scale is observed  at
a global bifurcation called  a 'saddle connection' \cite{Guckenheimer}.

In order to  present more concrete discussion,
we adapt [year] as  the time unit of the population dynamics
and set $\epsilon=10^{-6}$ as a typical value.
Then, our result implies that
(1) for  species shorter lived than  about $10^{7}$ [year]
    extinction occurs randomly with respect to the age of the species
    at a constant  rate of about $3 \times 10^{-7}$ [year$^{-1}$], and
(2) for species much longer lived  than about $ 10^{7}$  [year],
    older species  have a smaller probability of extinction.
The first result agrees with the taxonomic data of Van Valen \cite{Valen}.
Also, there are some taxonomic data consistent with our second result
\cite{Valen,Raup},
though we have not checked the power law distribution for the lifetimes  of
longer lived species.
According to our interpretation,
the time scale  of extinction is determined
by slowness of small perturbations  for  population dynamics.
We must consider what factors introduce the  most relevant perturbations
so as to estimate the  value of $\epsilon$.
Mutations of genes may be one candidate.
It would be interesting to study the population dynamics
of a system  taking into account the effect of mutations
\cite{Kauffmann,Lind,Kaneko}.


We thank  T. Yanagida, Y. Iba, K. Tokita, T. Ikegami and K. Kaneko
for fruitful discussions.  We also thank T. Shichijo and K. Watanabe
for their contributions to the early stage of this study.
G.C. Paquette is acknowledged for careful reading of the manuscript.
This work was supported in part by Foundation for Promotion of
Industrial Science. One of the authors (T. Chawanya)
acknowledges financial support from Yukawa foundation.



\vskip1cm


\begin{figure}
\caption{
Time development of populations of
$N=256$ species.
All populations within [0.01, 0.99] are plotted
every time unit.
An initial condition  was  chosen randomly.
An Euler method with a time mesh of 0.1 was adapted in the
numerical simulation.}
\label{fig1}
\end{figure}


\begin{figure}
\caption{
Plot of bursts in a species-$\tau$ plane after 10000 bursts
starting from an initial condition chosen randomly.}
\label{fig2}
\end{figure}


\begin{figure}
\caption{
Distribution of intervals between  two successive bursts
for $N=256$ species.
The distribution  is made from data of 40000 bursts after
discarding the initial transient behavior. 
The distribution curve is well-approximated
as $P(\tau)=\alpha \exp(- \alpha \tau)$, where $\alpha=25$.
$\alpha$ would increase linearly with $N$.}
\label{fig3}
\end{figure}


\begin{figure}
\caption{Distribution of persistence time of  positive $h_i$.
The distribution is made from data of 80000 bursts after
discarding the transient. 
The distribution curve has a power law tail for $\tau >\tau_c$, while
it is consistent with a Poisson distribution for $\tau <\tau_c$.
The cross-over time $\tau_c$ is about 10.  }
\label{fig4}
\end{figure}

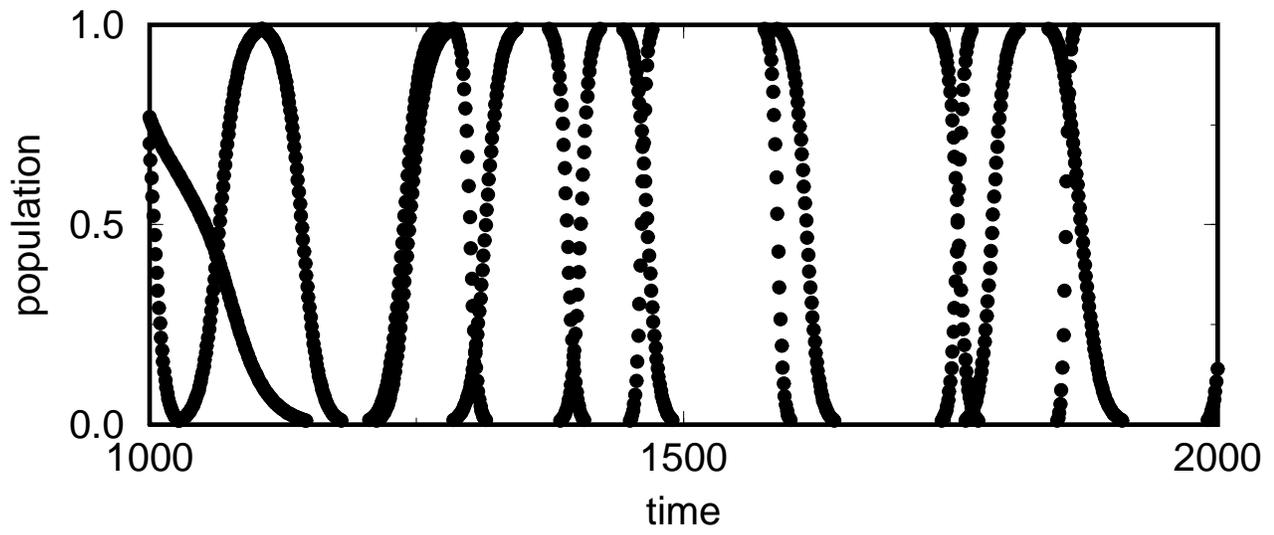
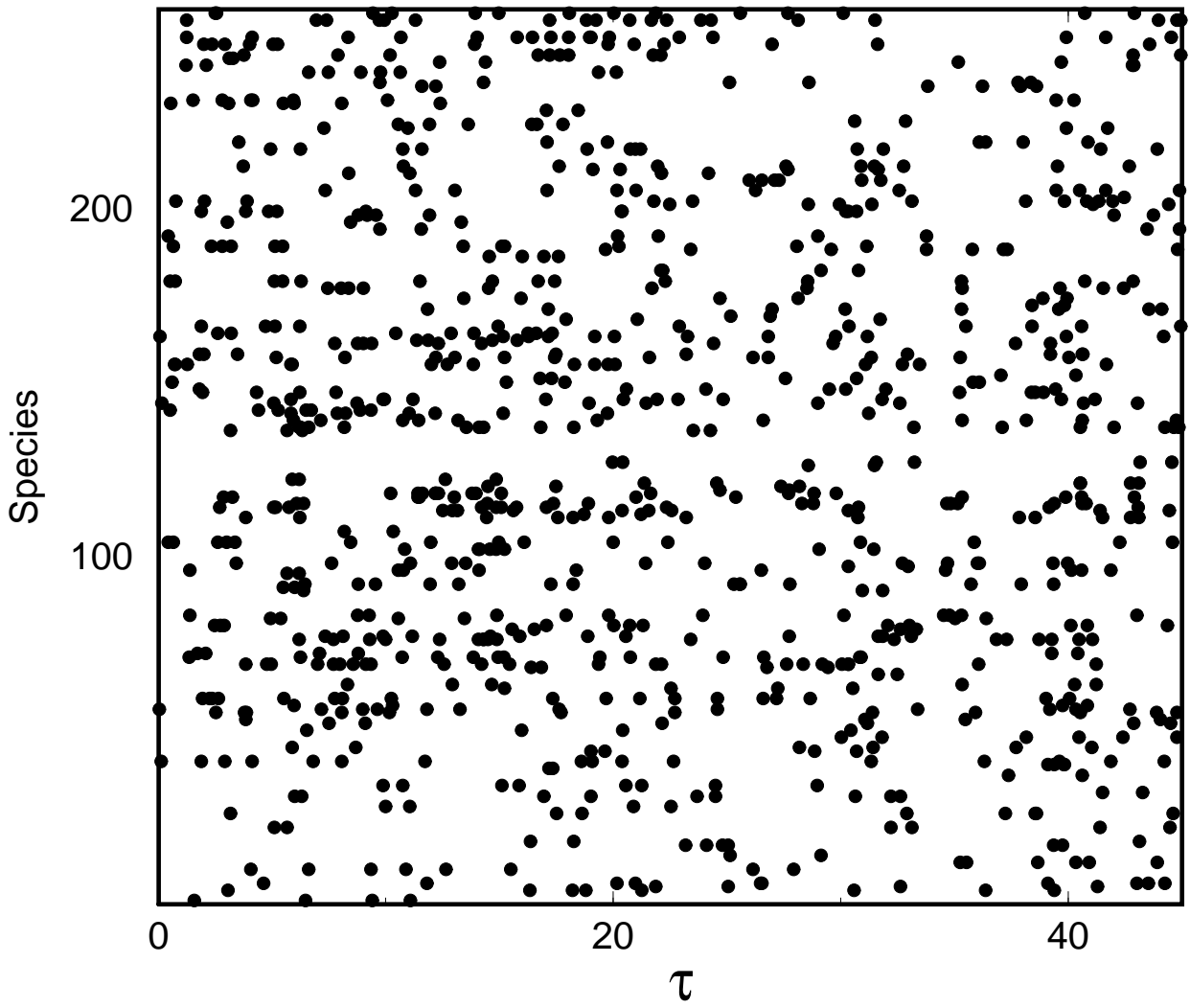
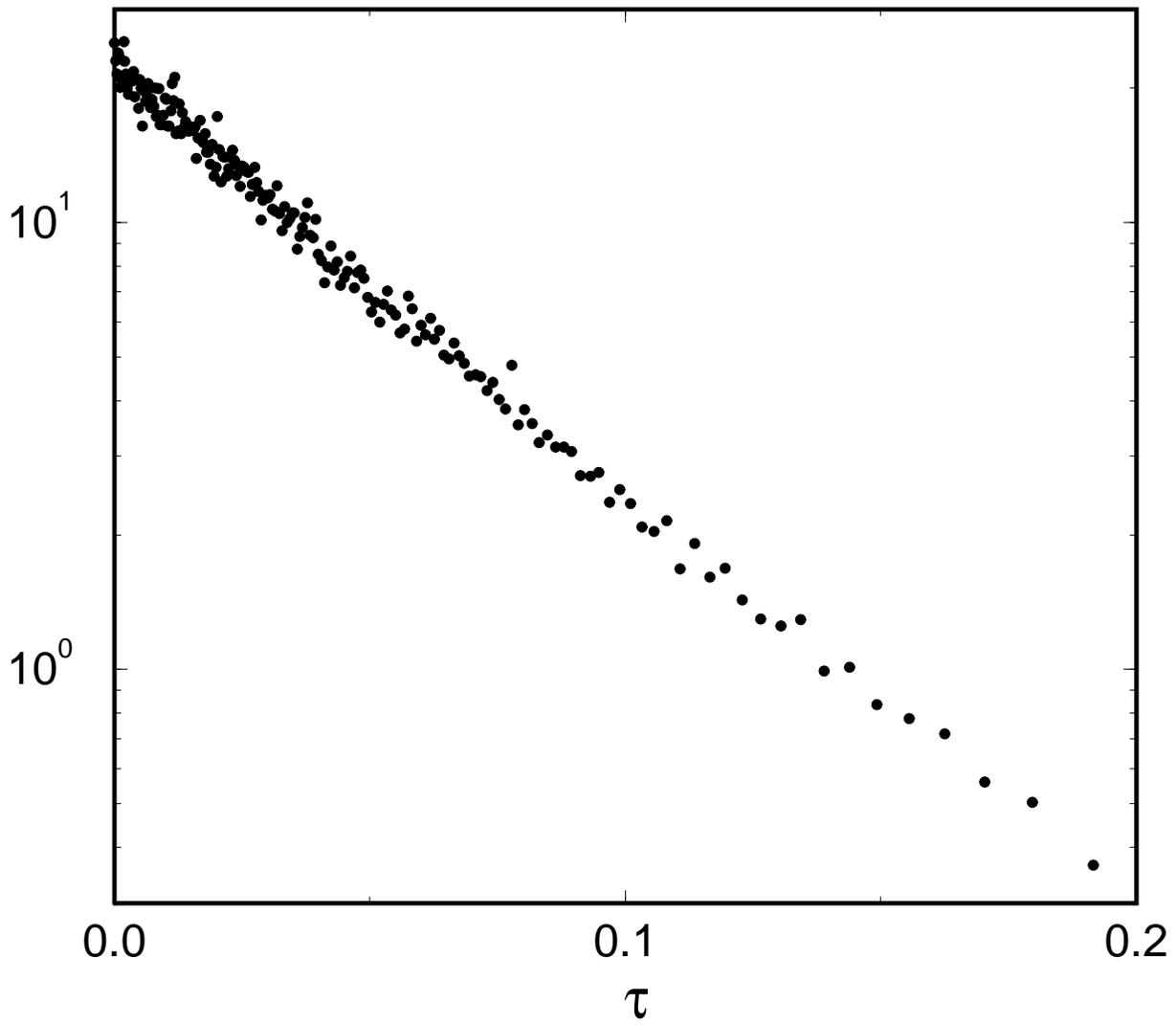
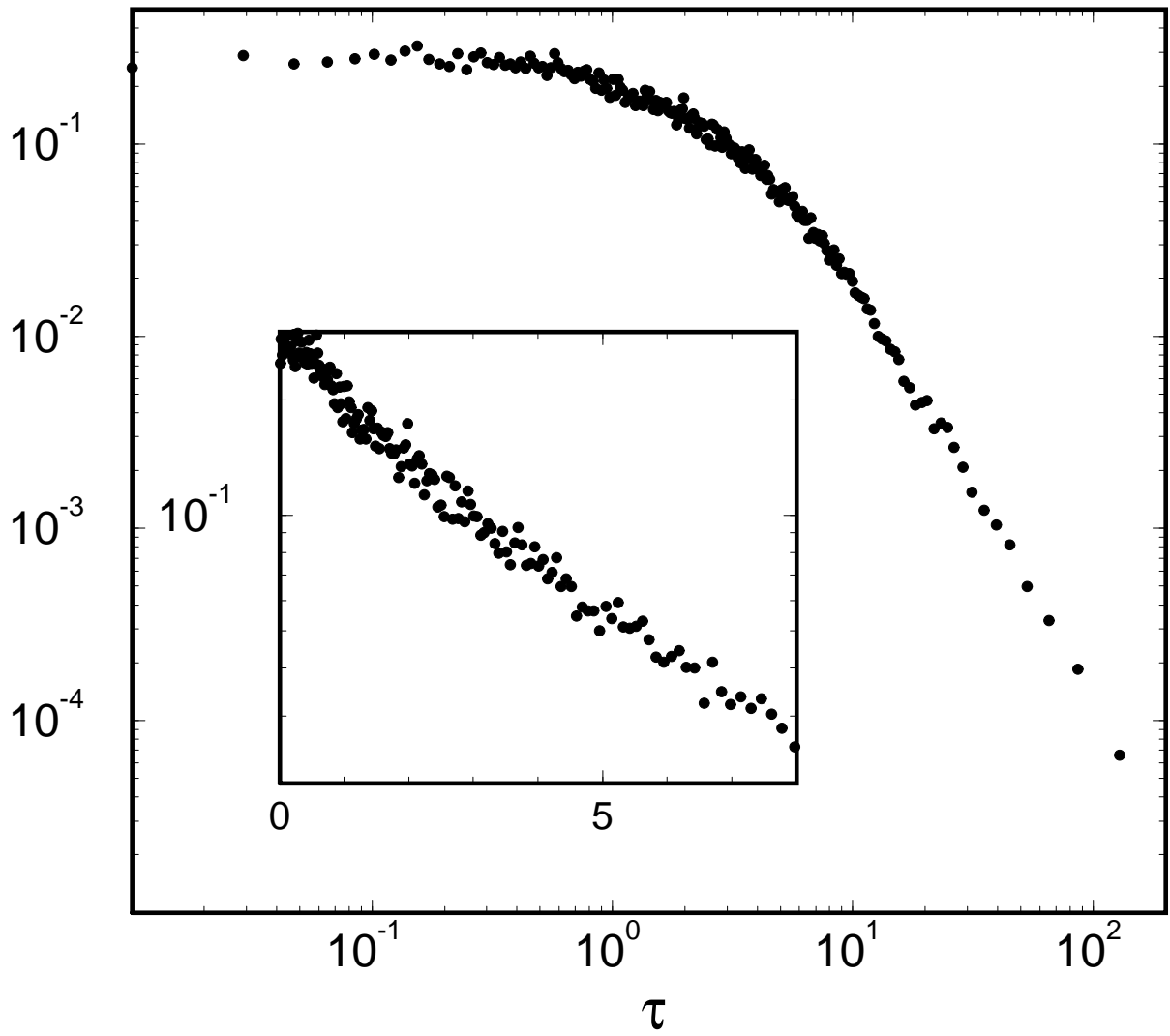
\end{document}